\documentclass[final,times,5p,twocolumn]{elsarticle}

\usepackage{graphicx}

\usepackage{amssymb}

\biboptions{sort&compress}

\usepackage{amsmath}

\graphicspath{{figures/}}

\journal{Optics Communications}

\begin{document}

\begin{frontmatter}



\title{Light scattering under nanofocusing: Towards coherent nanoscopies}


\author[label1]{Ahmad Mohammadi}
\author[label2]{Mario Agio}
\ead{mario.agio@phys.chem.ethz.ch}
\address[label1]{Department of Physics, Persian Gulf University, 75196 Bushehr, Iran}
\address[label2]{Laboratory of Physical Chemistry, ETH Zurich, 8093 Zurich, Switzerland}

\begin{abstract}
We investigate light scattering under nanofocusing in the context
of coherent spectroscopy. We discuss the different mechanisms
that may enhance the signal in extinction and how these depend on
nanofocusing as well as on the probed system.
We find that nanofocusing may improve the detection
sensitivity by orders of magnitude under realistic conditions,
enabling scanning implementations of coherent spectroscopy
at the nanoscale.
\end{abstract}


\end{frontmatter}


\section{Introduction}
\label{sec:introduction}

Optical nanoscopy is recognized as a
powerful tool for investigating physico-chemical
as well as biological processes in nanomaterials, whereby
subwavelength spatial resolution is combined
with a variety of spectroscopic techniques.
The past decades have witnessed the development
of several successful attempts in this regard,
such as scanning near-field optical microscopy
(SNOM)~\citep{pohl84,lewis84,inouye94,betzig92,hartschuh08},
single-molecule spectroscopy~\citep{orrit90,xie98,weiss99,moerner02,lupton10}
and the combination of the two~\citep{moerner94a,hess94,trautman94,betzig93}.
These efforts have however suffered from the mismatch between
light and nanoscale matter, which leads to a 
small throughput between the input and output signals.
That is why coherent detection schemes, whereby input
and output are in a well defined phase relationship, have found
less application, although they may provide additional
information~\citep{moerner89,kador99,zenhausern95,mikhailovsky03}. 

Recent developments in single-molecule spectroscopy
have however been able to push the limits of coherent
detection under various
conditions~\citep{gerhardt07a,wrigge08a,kukura10,celebrano11,pototschnig11}.
On the theoretical side, it has been found that there is a profound
relationship between the optimal concentration
of electromagnetic energy and the strength of light-matter
interaction~\citep{zumofen08,mojarad09b}, pointing out how
the possibility of focusing light below the diffraction limit
may enhance the detected signal.
The immediate question that arises is thus how to improve these
schemes further and how to implement them
for optical nanoscopies.

To address these questions we first note that the dramatic advances
of nanotechnology experienced in recent years have
enabled us to fabricate optical nanostructures that greatly improve
the conversion of localized electromagnetic energy into
radiation and vice versa~\citep{greffet05,muehlschlegel05,novotny11}.
In this context, single-molecule studies have shown how these so-called
optical antennas may provide an unprecedented control over
molecular-fluorescence~\citep{anger06,kuehn06a,kuehn08,taminiau08a,curto10}.

Recently, it has been found that a truncated metal nanocone
may focus light into nanoscale dimensions with minimal suffering
from absorption losses~\citep{chen10} and modify the
radiation pattern of a light emitter placed near its tip~\citep{chang06}. 
Furthermore, such antenna architecture is fully compatible with
scanning probe technology~\citep{deangelis10}. 
Nanofocusing relies on the propagation of surface plasmon polaritons
(SPPs) along a metal nanowire~\citep{takahara97}.
These modes do not have a cut-off and propagate optical energy
in a tapered structure up to the very end~\citep{babadjanyan00,stockman04}.
Note that this phenomenon applies to a variety of waveguide
geometries~\citep{gramotnev10}.

Our aim is to combine coherent spectroscopy with nanofocusing to explore
how this approach may expand the detection limits of nanoscale
objects, with particular attention on the competition between
the enhancement of light-matter interaction, coherent detection
as well as damping and dephasing processes. We furthermore highlight the unique features of
scattering under nanofocusing and derive expressions for the
visibility and the phase shift caused by a point-like 
polarizable object
placed in the near-field of the nanocone sharp end. We 
verify our ideas by means of rigorous electrodynamic calculations,
choosing very small metal particles to model nanoscale
emitters affected by strong non-radiative damping.

\section{Formulation}

\subsection{Coherent scattering under focusing}
\label{sec:focusing}

Textbook treatments of coherent scattering consider
a polarized plane wave incident on a material object.
Under these assumptions we may take advantage of the optical
theorem to relate the signal collected by 
an infinitesimal detector placed in the forward direction
to the total amount of power $P_\mathrm{ext}$ removed from the beam
by the obstacle~\citep{bohren83b}. 
If $P_\mathrm{inc}$ is the incident power,
the normalized detected signal reads
\begin{equation}
\label{eq:ot}
S=1-P_\mathrm{ext}/P_\mathrm{inc}=1-V,
\end{equation}
where $V$ stands for visibility. 
An established approach to enhance $V$ relies on focusing light.
However, because the optical theorem fails when the incident
wave is not plane~\citep{lock95b}, we need to depart from
Eq.~(\ref{eq:ot}) and consider the scattering problem
from first principles.

\begin{figure}[!htb]
\centering{
\includegraphics[width=8.25cm]{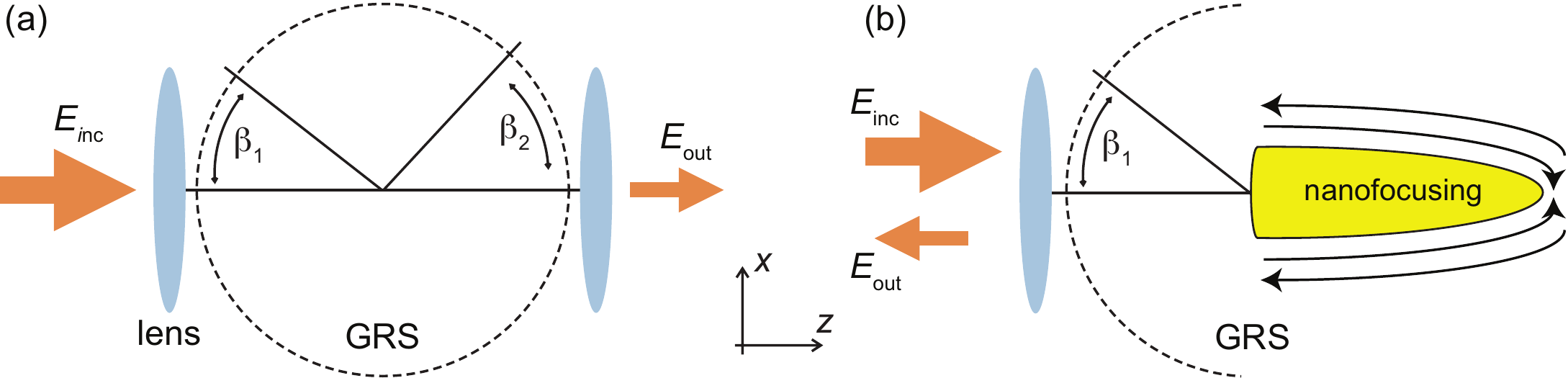}}
\caption{\label{fig:layout}
Coherent spectroscopy under focusing (a) and nanofocusing (b).
$\beta_1$ and $\beta_2$ are the focusing and collection semi-angles, respectively.
Note that in (b) focusing and collection occur through the same objective.
The dashed circle sketches the Gaussian reference sphere (GRS).
The black curves in (b) depict the propagation of SPPs in both directions.}
\end{figure}

Figure~\ref{fig:layout}a sketches the typical arrangement
of an extinction experiment where the incident field $E_\mathrm{inc}$
is focused by a lens and the output field $E_\mathrm{out}$
is collected by another lens placed in the forward direction.
According to scattering theory, $E_\mathrm{out}$ corresponds to the
sum of the incident and scattered fields.
Moreover, since we are interested in nanoscale objects,
we may say that
the scattered field $E_\mathrm{sca}$ is proportional to
the dipolar polarizability $\alpha$ and $E_\mathrm{inc}(O)$,
the incident field at the focal spot $O$.
For a classical oscillating dipole and a weakly excited
quantum emitter we find~\citep{jackson99,allen75}
\begin{equation}
\label{eq:pol-bare}
\alpha=-\dfrac{6\pi}{k^3}\dfrac{\Gamma_1}{2\Delta+i\Gamma_2},
\end{equation}
where $k$ is the wavevector, $\Delta$ is the detuning from resonance
and $\Gamma_1$, $\Gamma_2$ respectively are the radiative and total
damping rates. 

In free space there is a fundamental limit to the optimal concentration
of electromagnetic energy~\citep{bassett86}.
To understand how this gives rise
to an upper bound for $V$, we assume that $E_\mathrm{inc}$ is a
focused plane wave of amplitude $E_o$.
Furthermore, we explicitly consider the dependence on the
focusing and collection semi-angles.
In these circumstances Eq.~(\ref{eq:ot}) is replaced by
\begin{equation}
\label{eq:S-int}
S=1+\chi(\beta_2)P_\mathrm{sca}/P_\mathrm{inc}
+\int_0^{\Omega_{\beta_2}}2\mathrm{Re}
\{E_\mathrm{inc}E_\mathrm{sca}^*\}\mathrm{d}\Omega/P_\mathrm{inc},
\end{equation}
where $\mathrm{d}\Omega$ is the infinitesimal collection solid-angle,
whose maximum is $\Omega_{\beta_2}$, and 
$\chi(\beta_2)=\left(4-3\cos\beta_2-\cos^3\beta_2\right)/8$~\citep{zumofen08}.

If $\sigma_\mathrm{sca}=k^4|\alpha|^2/6\pi$ is the
scattering cross section~\citep{bohren83b}, the scattered power reads
\begin{equation}
\label{eq:Psca}
P_\mathrm{sca}=\sigma_\mathrm{sca}
\dfrac{k^2|E_o|^2f^2}{4}\mathcal{I}_o^2(\beta_1),
\end{equation}
where $f$ is the lens focal length and
$\mathcal{I}_o(\beta_1)$ is the diffraction integral
\begin{equation}
\mathcal{I}_o(\beta_1)=\int_0^{\beta_1}\mathrm{d}\theta
\sin\theta\sqrt{\cos\theta}(1+\cos\theta).
\end{equation}
The third term at the right side of Eq.~(\ref{eq:S-int})
represents the power $P_\mathrm{int}$ associated with the interference
between $E_\mathrm{inc}$ and $E_\mathrm{sca}$.
For $\beta_2>\beta_1$ we find
\begin{equation}
\label{eq:Pint}
P_\mathrm{int}=-\sigma_\mathrm{ext}\dfrac{k^2|E_o|^2f^2}{4}
\mathcal{I}_o^2(\beta_1).
\end{equation}
Note that $P_\mathrm{int}$ is negative and contains the so-called
extinction cross section
$\sigma_\mathrm{ext}=k\mathrm{Im}\{\alpha\}$~\citep{bohren83b}.
It corresponds to $P_\mathrm{ext}$ for plane-wave scattering.
Moreover, $|\mathcal{I}_o(\beta_1)|^2$ stems from two contributions.
One comes from $E_\mathrm{sca}$, since
$E_\mathrm{inc}(O)\propto\mathcal{I}_o(\beta_1)$. The other
one comes from the collection integral in Eq.~(\ref{eq:S-int}).
That explains how the visibility $P_\mathrm{int}/P_\mathrm{inc}$
builds up, showing that it depends twice on a diffraction integral:
first in the focusing process, second in the collection process.
Equations (\ref{eq:Psca}) and (\ref{eq:Pint}) are valid for
any type of focused beam if we replace
$k^2f^2|E_o|^2|\mathcal{I}_o(\beta_1)|^2/4$ with
$2cW_\mathrm{inc}(O)^\mathrm{(el)}$, where $c$ is the speed of light
and $W_\mathrm{inc}(O)^\mathrm{(el)}$ is the electric energy
density at the focal spot~\citep{zumofen08}.

\subsection{Visibility}

The maximum of $V$ is easily obtained starting from the
Bassett limit~\citep{bassett86}
\begin{equation}
W_\mathrm{inc}^\mathrm{(el)}(O)/P_\mathrm{inc}\le k^2/(6\pi c)
\end{equation}
and from the fact that an ideal oscillating dipole yields
$\sigma_\mathrm{ext}=\sigma_\mathrm{sca}=\sigma_o=6\pi/k^2$ for $\Delta=0$.
After replacing these quantities into Eqs.~(\ref{eq:Psca}) and
(\ref{eq:Pint}), Eq.~(\ref{eq:S-int}) yields $S=0$ and $V=1$.
Note that we have chosen $\beta_1=\beta_2=\pi/2$.
We furthermore point out that $V$ is directly related to the
phase shift $\phi$ that the scattering object induces in
$E_\mathrm{out}$~\citep{zumofen09,aljunid09}, namely
\begin{equation}
\phi=\arg(E_\mathrm{out}E_\mathrm{inc}^*)
=\arg\left(1+i\alpha
\dfrac{cW_\mathrm{inc}^\mathrm{(el)}(O)}{P_\mathrm{inc}}\right).
\end{equation}

We are interested in situations where the scatterer is far
from being ideal, i.e. where $\Gamma_2\gg \Gamma_1$.
To gain insight we set $\Delta=0$ and write
\begin{equation}
\label{eq:sigma-gamma}
\sigma_\mathrm{sca}=\sigma_o\left(\Gamma_1/\Gamma_2\right)^2,
\hspace{1cm}
\sigma_\mathrm{ext}=\sigma_o\left(\Gamma_1/\Gamma_2\right).
\end{equation}
Equation~(\ref{eq:sigma-gamma}) readily shows that
$P_\mathrm{sca}$ drops more rapidly than $P_\mathrm{int}$.
That is why detection of weakly scattering objects
is best achieved in extinction
experiments~\citep{kukura10,celebrano11}.

To analyze the visibility signal further, we choose a focused
radially-polarized beam (FRB),
because it allows a more direct comparison with nanofocusing.
We remind that although $V$ is maximal at the Bassett limit,
it is only a few percents smaller for a FRB~\citep{agio09}.
The  electric field of a radially-polarized beam
on the Gaussian reference sphere of an aplanatic
system reads
\begin{equation}
\label{eq:FRB}
E_\mathrm{inc}(a,\theta)=E_o\sqrt{\cos\theta}
\exp(-a^2\sin^2\theta/2)a\sin\theta,
\end{equation}
where $E_o$ is the field amplitude, $\theta$ is the azimuthal angle
with respect to the optical axis and $\sqrt{\cos\theta}$
is the apodization function. $a=f/w$ is the ratio between the
lens focal length $f$ and the beam waist $w$.

We refer to Eq.~(\ref{eq:S-int})
and consider full focusing
and collection in the forward direction, i.e. $\beta_1=\beta_2=\pi/2$.
A few algebraic steps lead to~\citep{agio09}
\begin{equation}
\label{eq:S-FRB}
S=1-\dfrac{6a^4\mathcal{I}_1^2(a,\pi/2)}{1-(1+a^2)\exp(-a^2)}
\dfrac{\Gamma_1}{\Gamma_2}\left(1-\dfrac{\Gamma_1}{2\Gamma_2}\right),
\end{equation}
where $\mathcal{I}_1(a,\beta_1)$ is the diffraction integral
\begin{equation}
\mathcal{I}_1(a,\beta_1)=\int_0^{\beta_1}\mathrm{d}\theta
\sin^3\theta\sqrt{\cos\theta}\exp
\left(-a^2\sin^2\theta/2\right).
\end{equation}

\begin{figure}[!htb]
\centering{
\includegraphics[width=8.25cm]{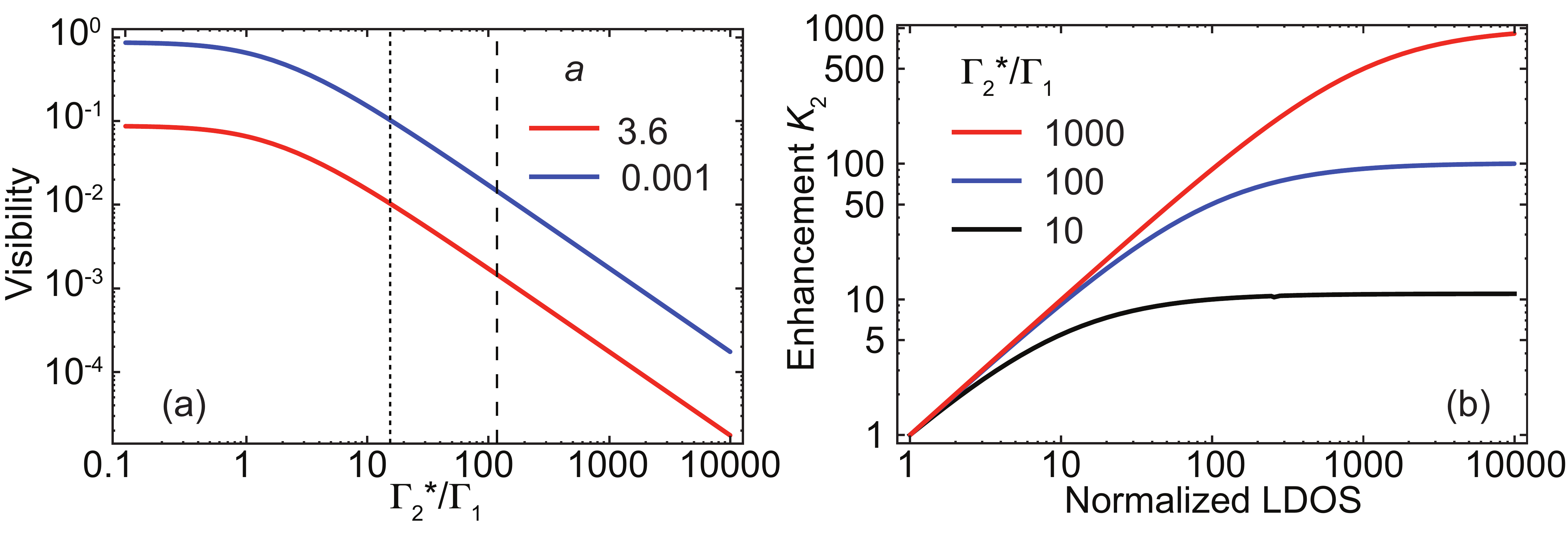}}
\caption{\label{fig:vis-quench}
(a) Visibility as a function of $\Gamma_2^*/\Gamma_1$ for two values
of the beam parameter $a$. For $a\to 0$ the FRB reaches its maximal
focusing strength. $a=3.6$ corresponds to the FRB used for nanofocusing.
The vertical dashed (dotted) line refers to the nanoparticle used in
Fig.~\ref{fig:S-nanofocusing}b (Fig.~\ref{fig:S-nanofocusing}c). 
(b) Visibility enhancement $K_2$ as a function of the normalized LDOS for different
values of $\Gamma_2^*/\Gamma_1$.}
\end{figure}

Figure~\ref{fig:vis-quench}a plots $V$ as a function
of $\Gamma_2^*/\Gamma_1$ for two values of the beam parameter
$a$, where $\Gamma_2=\Gamma_1+\Gamma_2^*$.
The curves exhibit a clear crossover between the region where
radiative and non-radiative damping dominate. 
In the first case $V$ is weakly dependent on $\Gamma_2^*$
and it is only determined by the focusing strength.
In the second case $V$ drops quite rapidly and even
the strongest focused beam yields a tiny signal.

By inspection of Eq.~(\ref{eq:S-FRB}) we argue that
$V$ may increase up to its maximum if
the radiative rate $\Gamma_1$ be enhanced at will.
Figure~\ref{fig:vis-quench}b illustrates that a
change in the local density of states (LDOS)
improves $V$ by a factor
\begin{equation}
\label{eq:K2}
K_2=\dfrac{F\Gamma_2}{F\Gamma_1+\Gamma_2^*},
\end{equation}
where $F$ represents the LDOS enhancement.
Note that Eq.~(\ref{eq:K2}) is valid either for a classical
oscillating dipole, where $\Gamma_1$ and $\Gamma_2^*$
refer to radiation and absorption, respectively,
either for a quantum emitter, where $\Gamma_1$ and $\Gamma_2^*$
represent spontaneous emission and non-radiative
decay~\citep{allen75}.
We furthermore identify a crossover
in $K_2$ when $F\Gamma_1$ becomes equal to $\Gamma_2^*$, whereby
for larger $F$ the visibility approaches
its maximum and $K_2$ saturates to $1+\Gamma_2^*/\Gamma_1$.

\subsection{Coherent scattering under nanofocusing}

We are now ready to explore how nanofocusing may provide
an effective way to achieve ultra-sensitive
coherent detection at the nanoscale
by combining large LDOS enhancements,
intense near fields and a nearly background-free
illumination and collection scheme.

We first recall how light propagates under
nanofocusing~\citep{stockman04,chang06}. 
As sketched in Fig.~\ref{fig:layout}b, both $E_\mathrm{inc}$
and $E_\mathrm{out}$ break the diffraction limit, because
the nanocone funnels these fields to and from the probed object
via the tip. A crucial feature of 
the configuration whereby a semi-infinite nanocone is replaced
by a conical antenna is the efficient conversion between
photons and SPPs. That is possible if we take a FRB and tune
$a$ (see Eq.~(\ref{eq:FRB})) such that the beam matches
the nanocone radiation pattern.
For the same purpose, the nanocone base diameter
needs to be accordingly chosen~\citep{chen10}.
When $E_\mathrm{inc}$ reaches the tip,
the nanocone radius is so small that
most of $P_\mathrm{inc}$ is reflected backwards~\citep{gordon09}
and it propagates towards the base, where it is radiated
into the far field.
Moreover, if a scatterer is placed near the tip,
$E_\mathrm{sca}$ couples to
the nanocone and reaches the lens in the same spatial
mode of the back-reflected $E_\mathrm{inc}$.

Our goal is to find the detected signal $S$, i.e.
the visibility $V$, and also the phase shift $\phi$ under nanofocusing.
In this derivation we assume that $E_\mathrm{inc}$ and $E_\mathrm{out}$
are perfectly converted into SPPs and vice versa, respectively.
We first consider $E_\mathrm{out}$ without scatterer near the
tip, as shown in Fig.~\ref{fig:layout}b.
In such case $E_\mathrm{out}=\kappa E_\mathrm{inc}$, where $|\kappa|<1$
accounts for scattering and absorption losses that occur
in the nanofocusing process~\citep{vogel07,gramotnev08}.

We now add a polarizable nanoscale object near the tip.
The calculation of $E_\mathrm{sca}$ requires some attention.
First, in the nanofocusing process $E_\mathrm{inc}$ is enhanced
by a factor $\sqrt{|\kappa|}\xi$ with respect the value at $O$
in free space. Here $\sqrt{|\kappa|}$ takes into account
the effect of scattering and absorption losses in reducing the
field enhancement.
Second, the fraction of $P_\mathrm{sca}$ that is coupled into the nanocone
is proportional to the so-called $\beta$ factor~\citep{chang06}.
Third, the polarizability is modified by the presence of the
tip, which enhances radiation damping and leads to
\begin{equation}
\label{eq:pol-nano}
\alpha=-\dfrac{6\pi}{k^3}\dfrac{1}{F}\dfrac{\Gamma_1'}{2\Delta+i\Gamma_2'}.
\end{equation}
Here the primes indicate modified rates according to
$\Gamma_1'=F\Gamma_1$ and $\Gamma_2'=F\Gamma_1+\Gamma_2^*$.
Fourth, due to absorption by real metals, the fraction of
$P_\mathrm{sca}$ that reaches the lens depends on $\eta_\mathrm{a}$,
the antenna efficiency~\citep{rogobete07a}.
Furthermore, for $\Gamma_2^*\gg \Gamma_1$
we may neglect the contribution of
$\sigma_\mathrm{sca}$ (see Eq.~(\ref{eq:sigma-gamma})).
Normalizing the signal by $|\kappa|^2 P_\mathrm{inc}$ leads to
\begin{equation}
\label{eq:S-nano}
S\simeq 1-\beta\dfrac{\eta_a}{|\kappa|}\sigma_\mathrm{ext}|\xi|^2
\dfrac{2cW_\mathrm{inc}^\mathrm{(el)}(O)}{P_\mathrm{inc}},
\end{equation}
where $W_\mathrm{inc}^\mathrm{(el)}(O)$ has been defined
in Sec.~\ref{sec:focusing}. A further simplification
stems from the fact that
$\eta_\mathrm{a}\simeq |\kappa|$. Hence, $V$ reads
\begin{equation}
\label{eq:vis-nano}
V\simeq \sigma_o\dfrac{\beta|\xi|^2}{F}
\dfrac{F\Gamma_1}{F\Gamma_1+\Gamma_2^*}
\dfrac{2cW_\mathrm{inc}^\mathrm{(el)}(O)}{P_\mathrm{inc}}
=V_o\dfrac{\beta|\xi|^2}{F}
\dfrac{F\Gamma_2}{F\Gamma_1+\Gamma_2^*},
\end{equation}
where $V_o$ corresponds to $V$ under focused illumination.
Recalling that $\beta=(F-1)/F$, the enhancement $K$
may be written as
\begin{equation}
\label{eq:K}
K=K_1K_2=
\underbrace{\dfrac{|\xi|^2(F-1)}{F^2}}_{K_1}
\underbrace{\dfrac{F\Gamma_2}{F\Gamma_1+\Gamma_2^*}}_{K_2}
\end{equation}
A similar expression is obtained when $\Gamma_2^*$
is due to elastic processes (dephasing).
Equation~(\ref{eq:K}) reveals that $V$ improves by two
mechanisms. Besides the one discussed in the previous section,
which we named $K_2$, we note an additional term $K_1$
that depends on $|\xi|^2$ and $F$.
For $F\gg 1$, $\beta$ is close to one and $K_1$ reduces
to $|\xi|^2/F$. While for resonant dipole antennas
it is common to find $|\xi|^2<F$, for nanofocusing 
we may have the opposite result. That is possible
because the incident beam is weakly focused
(see Fig.~\ref{fig:vis-quench}a).

At last, we wish to examine the role of $\kappa$ and $\eta_\mathrm{a}$.
Although they do not appear in Eq.~(\ref{eq:vis-nano}), they
are nevertheless important as the output signal is proportional
to $|\kappa|^2$ and $|\kappa|\eta_\mathrm{a}$.
Indeed, if $\kappa$ tends to zero the amount
of power that reaches the tip becomes negligible. 
Likewise if $\eta_\mathrm{a}$ is very small most of $P_\mathrm{sca}$
gets absorbed in the nanocone.
Thus the whole enhancement process is
meaningless, because it is given by
the division of two vanishing signals, namely $|\kappa|^2P_\mathrm{inc}$
and $\eta_\mathrm{a}|\kappa|P_\mathrm{sca}$. It is therefore crucial to
operate with antenna architectures that combine huge LDOS enhancements
with  moderate absorption losses.

\section{Results}

\subsection{Metal nanocones}

\begin{figure}[!htb]
\centering{
\includegraphics[width=6.5cm]{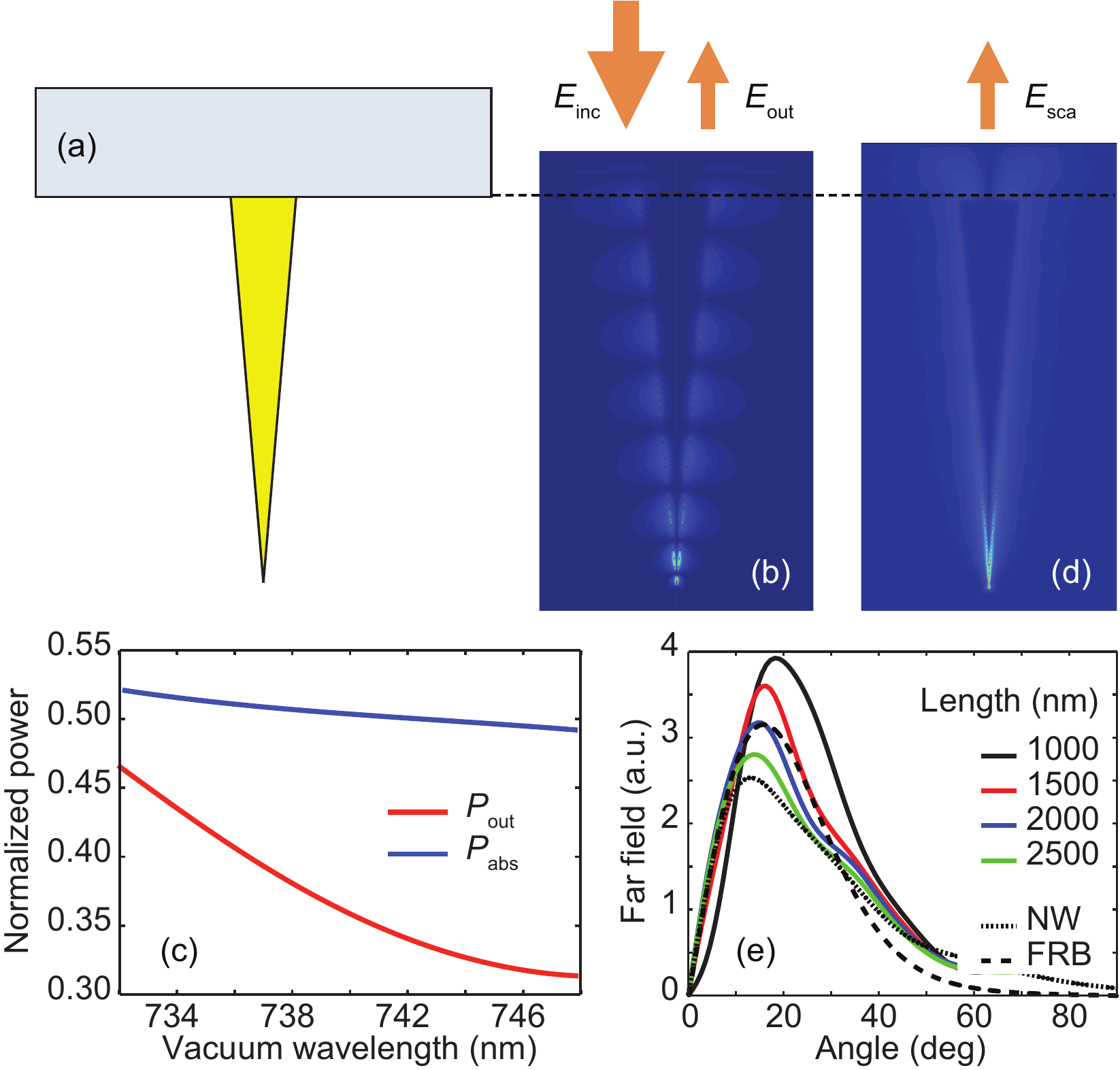}}
\caption{\label{fig:nanofocusing}
(a) Sketch of a nanocone mounted on a quartz AFM cantilever.
The nanocone has a base radius of 195 nm and a tip radius of 5 nm.
Its height is 2000 nm.
(b) $E_\mathrm{inc}$ and $E_\mathrm{out}$ without sample near the tip. 
(c) $P_\mathrm{out}$ and $P_\mathrm{abs}$
normalized with respect to $P_\mathrm{inc}$.
(d) $E_\mathrm{sca}$ generated by an oscillating dipole placed 10 nm from the tip.
In (b) and (d) the field patterns represent the magnetic field
amplitude at 740 nm.
(e) Far field as a function of the nanocone length at 740 nm.
The dashed and dotted curves respectively refer to the far field
of a FRB with $a=3.6$ and a gold nanowire (NW).
The base and tip radii are like in (a) for all lengths.}
\end{figure}

We have modeled gold nanocones using a body-of-revolution (BOR) finite-difference
time-domain (FDTD) approach~\citep{taflove05}.
Figure~\ref{fig:nanofocusing}a sketches a nanocone attached
to an AFM cantilever, treated as a semi-infinite quartz substrate.
We have considered nanocone lengths between 1000 and 3000 nm, and base diameters
optimized according to the wavelength range of interest~\citep{chen10}.
Without loss of generality we choose to work around 740 nm,
which corresponds to an optimal base radius of 195 nm.
The tip has a radius of curvature of 5 nm, which is
a realistic value for state of the art nanofabrication~\citep{deangelis10}.
If not otherwise stated, the FDTD mesh has a discretization of 1 nm.

We now move to the input and output fields. We illuminate the nanocone
using a FRB with $a=3.6$ to optimize the conversion of photons into
SPPs. Figure~\ref{fig:nanofocusing}b shows the magnetic
field amplitude as the wave propagates towards the tip and comes back.
A standing-wave pattern is indeed clearly visible. 
Moreover, note that its structure
changes as the field approaches the tip, because
the effective wavelength becomes shorter in the nanofocusing
process~\citep{babadjanyan00,stockman04}.
We have calculated how much power is absorbed by a 2000 nm long gold
nanocone, finding that
it amounts to nearly 50\% of $P_\mathrm{inc}$, hence $|\kappa|^2=0.5$.
Figure~\ref{fig:nanofocusing}c plots the absorbed power $P_\mathrm{abs}$
and the output power $P_\mathrm{out}$ as a function of wavelength.
Note that $P_\mathrm{out}$ is in fact a little less than
$P_\mathrm{inc}-P_\mathrm{abs}$,
because scattering losses occur during
nanofocusing and conversion of SPPs.

Next we introduce $E_\mathrm{sca}$ by placing an oscillating
dipole 10 nm from the tip, oriented along the optical
axis. Figure~\ref{fig:nanofocusing}d confirms that most of the power
is radiated along the nanocone and exits the other end with
an efficiency $\eta_\mathrm{a}$ of about 70\%,
which is consistent with $\eta_\mathrm{a}\simeq |\kappa|$. 
To explore how $E_\mathrm{sca}$ behaves in the far field,
we have performed a near-to-far field transformation. The result is plotted
in Fig.~\ref{fig:nanofocusing}e for different nanocone lengths.
As the nanocone gets longer the far field approaches that of a metal nanowire,
indicated by a dotted curve. The dashed curve shows instead the
profile of a FRB for $a=3.6$. We find that $a$ needs
to be adjusted to optimize the coupling also in relation to the
nanocone length. In practice, a shorter nanocone requires a FRB with a slightly
smaller $a$. Note that the length is an important parameter
as it determines nanofocusing strength and propagation losses, which
also depend on the nanocone composition and working
wavelength~\citep{vogel07,gramotnev08}.

\begin{figure}[!htb]
\centering{
\includegraphics[width=8.25cm]{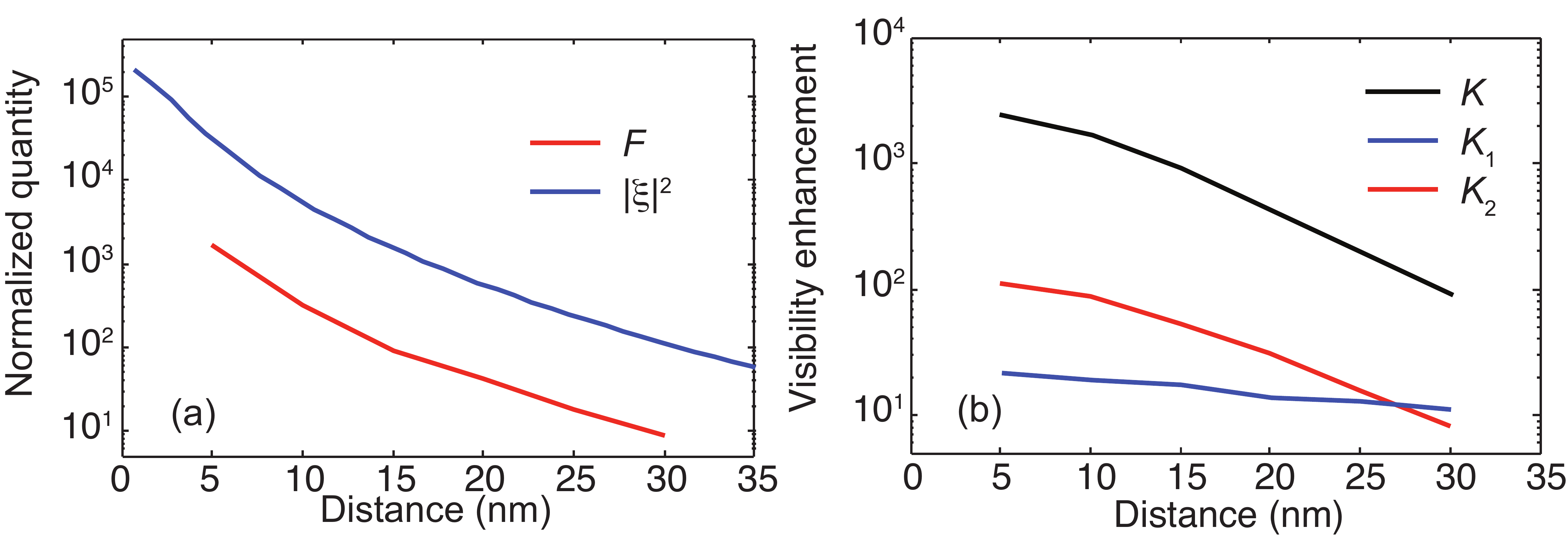}}
\caption{\label{fig:eff-nanofocusing}
(a) Intensity enhancement $|\xi|^2$, normalized LDOS $F$ and
(b) visibility enhancement $K$ as a function of distance.
The individual contributions $K_1$ and $K_2$ are also indicated.
For the beam and nanocone parameters see Fig.~\ref{fig:nanofocusing}.
$\Gamma_2^*/\Gamma_1$ is equal to 119, which corresponds to
the nanoparticle discussed in Fig.~\ref{fig:S-nanofocusing}b.}
\end{figure}

Having analyzed the behavior of the input and output fields, we move
our attention to the enhancement factors. These are plotted in
Fig.~\ref{fig:eff-nanofocusing}a, which shows $|\xi|^2$ and $F$
as a function of distance from the tip.
Both intensity and LDOS enhancements are strongly distance dependent
and grow very rapidly as we approach the tip.

We are ready to quantify the visibility enhancements $K_1$ and $K_2$, 
which are shown Fig.~\ref{fig:eff-nanofocusing}b for
$\Gamma_2^*/\Gamma_1\simeq 100$. We remark that $K_1$ is independent of
$\Gamma_2^*/\Gamma_1$ as it contains only $|\xi|^2$ and $F$.
Furthermore, it weakly depends on the distance from the tip.
On the other hand, $K_2$ changes by more than one order of magnitude
as the distance is reduced by a few tens of nanometers.
Moreover, it approaches a plateau for distances less
than 10 nm. There, $F$ becomes larger than $\Gamma_2^*/\Gamma_1$
and $K_2$ saturates to its maximum
(see Fig.~\ref{fig:vis-quench}b). The cooperation
of $K_1$ and $K_2$ leads to an overall enhancement of more than
three orders of magnitude.

\subsection{Metal nanoparticles}

We verify our findings by performing coherent 
spectroscopy on very small metal particles.
To gain insight we treat them
as a point-like oscillating dipole with radiative 
corrections~\citep{wokaun82}, whose polarizability reads
\begin{equation} \label{eq:pol-NP} 
\alpha\simeq -4\pi r^3\dfrac{\omega_o}{2\Delta+i(\gamma+2\omega_o(kr)^3/3)}, 
\end{equation}
where $r$ is the nanoparticle radius.
This expression is analogous to Eq.~(\ref{eq:pol-bare}).
We have used the Drude model
to describe the nanosphere dielectric function
$\epsilon(\omega)=1-\omega_\mathrm{p}^2/\omega(\omega+i\gamma)$, 
where $\omega_\mathrm{p}$ and $\gamma$ are the so-called plasma and damping 
frequencies~\citep{ashcroft76}. In Eq.~(\ref{eq:pol-NP}) 
$\omega_o$ stands for $\omega_\mathrm{p}/\sqrt{3}$.

$\Gamma_2^*$ represents material losses 
in the metal nanosphere and it corresponds to $\gamma$.
Radiation losses are instead parametrized by $(kr)^3$.
A few algebraic steps lead to
\begin{equation}
\dfrac{\Gamma_2^*}{\Gamma_1}=\dfrac{\gamma}{\omega_\mathrm{p}}
\dfrac{3\sqrt{3}}{2(kr)^3}.
\end{equation}
Although $\gamma/\omega_\mathrm{p}$ is commonly much smaller than
one, the possibility that $kr\ll 1$ may lead
to substantial quenching of radiative damping and to a considerable
increase of $\Gamma_2^*/\Gamma_1$.

The strength of light-matter interaction gets smaller due to a decrease in
the polarizability (see Eq.~(\ref{eq:pol-NP})).
Moreover, the nanoparticle absorbs most of the light that
is coupled to it. To study how these phenomena affect $V$, 
we respectively introduce the nanoparticle and the scattering
efficiencies~\citep{agio09},
\begin{equation}
\eta_\mathrm{NP} = \dfrac{\sigma_\mathrm{ext}}{\sigma_o}=
\dfrac{k^3\mathrm{Im}(\alpha)}{6\pi},
\hspace{1cm}
\eta_\mathrm{sca} = \dfrac{\sigma_\mathrm{sca}}{\sigma_\mathrm{ext}}=
\dfrac{k^3|\alpha|^2}{6\pi\mathrm{Im}(\alpha)},
\end{equation}
which turn out to be 
\begin{equation}
\eta_\mathrm{NP}=\eta_\mathrm{sca}=
\left(1+\dfrac{\Gamma_2^*}{\Gamma_1}\right)^{-1}.
\end{equation}
For our analysis we set $\gamma/\omega_\mathrm{p}=0.001$
and two different nanoparticle radii, namely 2.5 and 5 nm.
Moreover, we choose $\omega_\mathrm{p}$ such that the
resonance wavelength is 675 nm and 740 nm,
respectively. These parameters lead to
$\Gamma_2^*/\Gamma_1\simeq 119$ for $r=2.5$ nm and
$\Gamma_2^*/\Gamma_1\simeq 16$ for $r=5$ nm.
The corresponding $V$ under focusing is
marked by vertical lines in Fig.~\ref{fig:vis-quench}a.
It amounts to 1-10\% for the strongest focused beam and
to 0.1-1\% for the FRB used for nanofocusing.

\subsection{Visibility and phase shift under nanofocusing}

\begin{figure}
\centering{
\includegraphics[width=8.25cm]{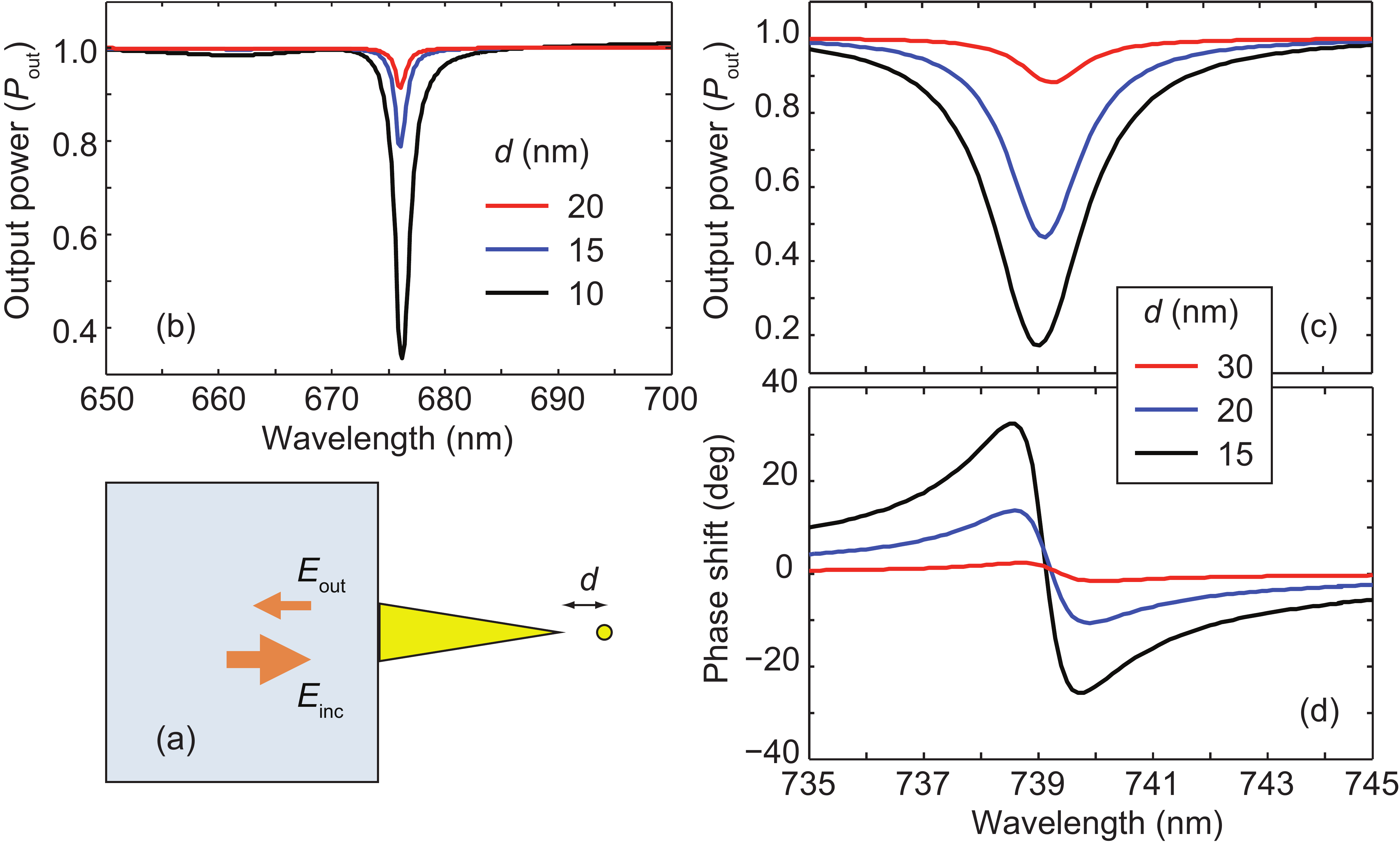}}
\caption{\label{fig:S-nanofocusing}
(a) Layout of the scattering problem (nanoparticle not to scale).
Signal ((b) and (c)) and phase shift (d) as a function of
wavelength for various distances from the tip.
The nanoparticle radius is 2.5 nm (b) or 5 nm ((c) and (d))
(see text for details).}
\end{figure}

To facilitate the FDTD calculations we have reduced the nanocone length
to 1000 nm and increased the mesh pitch to 2 nm.
While the latter may reduce the accuracy of our results, it does
not fake the main effect that we are interested in, namely the
enhancement of $V$. The layout of the problem is sketched in
Fig.~\ref{fig:S-nanofocusing}a.

Figures~\ref{fig:S-nanofocusing}b and \ref{fig:S-nanofocusing}c
show that $V$ up to 83\%
may be achieved if the nanoparticle is probed by a nanofocused beam.
In comparison to the curve in Fig.~\ref{fig:vis-quench}a,
it corresponds to enhancements of more than two orders of magnitude.
As expected, to achieve the same signal with the
smaller object we need to operate at shorter distances.
As the nanocone approaches
the nanosphere we recognize an increase in radiative damping
by broadening of the dip in $P_\mathrm{out}$.
Note that $P_\mathrm{out}$ without nanoparticle is about 44\%,
in agreement with the result of Fig.~\ref{fig:nanofocusing}d,
which refers to a nanocone length of 2000 nm.
The enhancement of light-matter interaction is noticeable in
the phase shift too, which is plotted in Fig.~\ref{fig:S-nanofocusing}d.
We finally remark that $K$ may be further improved
by optimizing the nanocone parameters~\citep{vogel07,gramotnev08},
reducing the working distance and choosing systems
with larger $\Gamma_2^*/\Gamma_1$.

\section{Conclusions}

We have proposed an implementation of coherent optical nanoscopy
based on the nanofocusing concept and on scanning-probe technology.
We have discussed the basic principles of this method and
its potential in enhancing the detection of nanoscale objects.
We have identified the two mechanisms that increase $V$,
namely the intensity enhancement and the modification of the LDOS
at the nanocone sharp end. While the second effect has been first
investigated in single-molecule SNOM~\citep{ambrose94,bian95}
and more recently in quantum-optics~\citep{chang06},
here we have shown how it may
increase the visibility and the phase shift caused by
an oscillating dipole.

In comparison with grating coupled nanofocusing~\citep{ropers07},
which also provides a nanoscale source for scattering-type
SNOM~\citep{neacsu10,sadiq11}, our approach exploits nanofocusing
in both illumination and collection channels and realizes
a kind of $4\pi$ optical system.
That is not easy to implement with grating couplers,
because the throughput may not be high enough.

In comparison with resonant optical antennas, whereby coherent
coupling has been studied from a quantum-optical
perspective~\citep{ridolfo10}, our scheme does not suffer
from resonance shifts that occur when a nano-emitter
approaches the antenna~\citep{alaverdyan11}.
We remark that these may completely detuned the antenna
and compromise the enhancements.

We wish to emphasize that these concepts are readily extendible
to ultrafast and nonlinear
techniques~\citep{sanchez99,guenther02,ichimura04,celebrano09,berweger11},
as well as coherent control~\citep{brinks10},
whereby short laser pulses are efficiently converted into SPPs
and nanofocused.
Multidimensional correlation spectroscopies~\citep{mukamel00}
may also be implemented at the nanoscale
in this way, without the need for hybrid approaches~\citep{aeschlimann11}.
Lastly, we wish to point out that these ideas may also be
pursued in the infrared spectral range
using surface phonon-polaritons~\citep{hillenbrand02} or at THz
and lower frequencies using spoof SPPs~\citep{maier06a,goubau50}.

\section*{Acknowledgments}

A. Mohammadi is thankful to the Persian Gulf University
Research Council for continuous support.
M. Agio wish to thank Vahid Sandoghdar for advice
and encouragement and Xuewen Chen for fruitful discussions.





\bibliographystyle{model1-num-names}

\begin{thebibliography}{73}
\expandafter\ifx\csname natexlab\endcsname\relax\def\natexlab#1{#1}\fi
\providecommand{\bibinfo}[2]{#2}
\ifx\xfnm\relax \def\xfnm[#1]{\unskip,\space#1}\fi
\bibitem[{Pohl et~al.(1984)Pohl, Denk, and Lanz}]{pohl84}
\bibinfo{author}{D.~W. Pohl}, \bibinfo{author}{W.~Denk},
  \bibinfo{author}{M.~Lanz},
\newblock \bibinfo{journal}{Appl. Phys. Lett.} \bibinfo{volume}{44}
  (\bibinfo{year}{1984}) \bibinfo{pages}{651--653}.
\bibitem[{Lewis et~al.(1984)Lewis, Isaacson, Harootunian, and Muray}]{lewis84}
\bibinfo{author}{A.~Lewis}, \bibinfo{author}{M.~Isaacson},
  \bibinfo{author}{A.~Harootunian}, \bibinfo{author}{A.~Muray},
\newblock \bibinfo{journal}{Ultramicroscopy} \bibinfo{volume}{13}
  (\bibinfo{year}{1984}) \bibinfo{pages}{227--231}.
\bibitem[{Inouye and Kawata(1994)}]{inouye94}
\bibinfo{author}{Y.~Inouye}, \bibinfo{author}{S.~Kawata},
\newblock \bibinfo{journal}{Opt. Lett.} \bibinfo{volume}{19}
  (\bibinfo{year}{1994}) \bibinfo{pages}{159--161}.
\bibitem[{Hartschuh(2008)}]{hartschuh08}
\bibinfo{author}{A.~Hartschuh},
\newblock \bibinfo{journal}{Angew. Chem. Int. Ed.} \bibinfo{volume}{47}
  (\bibinfo{year}{2008}) \bibinfo{pages}{8178--8191}.
\bibitem[{Betzig and Trautman(1992)}]{betzig92}
\bibinfo{author}{E.~Betzig}, \bibinfo{author}{J.~K. Trautman},
\newblock \bibinfo{journal}{Science} \bibinfo{volume}{257}
  (\bibinfo{year}{1992}) \bibinfo{pages}{189--195}.
\bibitem[{Xie and Trautman(1998)}]{xie98}
\bibinfo{author}{X.~S. Xie}, \bibinfo{author}{J.~K. Trautman},
\newblock \bibinfo{journal}{Annu. Rev. Phys. Chem.} \bibinfo{volume}{49}
  (\bibinfo{year}{1998}) \bibinfo{pages}{441--480}.
\bibitem[{Weiss(1999)}]{weiss99}
\bibinfo{author}{S.~Weiss},
\newblock \bibinfo{journal}{Science} \bibinfo{volume}{283}
  (\bibinfo{year}{1999}) \bibinfo{pages}{1676--1683}.
\bibitem[{Moerner(2002)}]{moerner02}
\bibinfo{author}{W.~E. Moerner},
\newblock \bibinfo{journal}{J. Phys. Chem. B} \bibinfo{volume}{106}
  (\bibinfo{year}{2002}) \bibinfo{pages}{910–927}.
\bibitem[{Lupton(2010)}]{lupton10}
\bibinfo{author}{J.~M. Lupton},
\newblock \bibinfo{journal}{Adv. Mater.} \bibinfo{volume}{22}
  (\bibinfo{year}{2010}) \bibinfo{pages}{1689--1721}.
\bibitem[{Orrit and Bernard(1990)}]{orrit90}
\bibinfo{author}{M.~Orrit}, \bibinfo{author}{J.~Bernard},
\newblock \bibinfo{journal}{Phys. Rev. Lett.} \bibinfo{volume}{65}
  (\bibinfo{year}{1990}) \bibinfo{pages}{2716--2719}.
\bibitem[{Moerner et~al.(1994)Moerner, Plakhotnik, Irngartinger, Wild, Pohl,
  and Hecht}]{moerner94a}
\bibinfo{author}{W.~E. Moerner}, \bibinfo{author}{T.~Plakhotnik},
  \bibinfo{author}{T.~Irngartinger}, \bibinfo{author}{U.~P. Wild},
  \bibinfo{author}{D.~W. Pohl}, \bibinfo{author}{B.~Hecht},
\newblock \bibinfo{journal}{Phys. Rev. Lett.} \bibinfo{volume}{73}
  (\bibinfo{year}{1994}) \bibinfo{pages}{2764--2767}.
\bibitem[{Hess et~al.(1994)Hess, Betzig, Harris, Pfeiffer, and West}]{hess94}
\bibinfo{author}{H.~F. Hess}, \bibinfo{author}{E.~Betzig},
  \bibinfo{author}{T.~D. Harris}, \bibinfo{author}{L.~N. Pfeiffer},
  \bibinfo{author}{K.~W. West},
\newblock \bibinfo{journal}{Science} \bibinfo{volume}{264}
  (\bibinfo{year}{1994}) \bibinfo{pages}{1740--1745}.
\bibitem[{Trautman et~al.(1994)Trautman, Macklin, Brus, and
  Betzig}]{trautman94}
\bibinfo{author}{J.~K. Trautman}, \bibinfo{author}{J.~J. Macklin},
  \bibinfo{author}{L.~E. Brus}, \bibinfo{author}{E.~Betzig},
\newblock \bibinfo{journal}{Nature} \bibinfo{volume}{369}
  (\bibinfo{year}{1994}) \bibinfo{pages}{40--42}.
\bibitem[{Betzig and Chichester(1993)}]{betzig93}
\bibinfo{author}{E.~Betzig}, \bibinfo{author}{R.~J. Chichester},
\newblock \bibinfo{journal}{Science} \bibinfo{volume}{262}
  (\bibinfo{year}{1993}) \bibinfo{pages}{1422--1425}.
\bibitem[{Mikhailovsky et~al.(2003)Mikhailovsky, Petruska, Stockman, and
  Klimov}]{mikhailovsky03}
\bibinfo{author}{A.~A. Mikhailovsky}, \bibinfo{author}{M.~A. Petruska},
  \bibinfo{author}{M.~I. Stockman}, \bibinfo{author}{V.~I. Klimov},
\newblock \bibinfo{journal}{Opt. Lett.} \bibinfo{volume}{28}
  (\bibinfo{year}{2003}) \bibinfo{pages}{1686--1688}.
\bibitem[{Moerner and Kador(1989)}]{moerner89}
\bibinfo{author}{W.~E. Moerner}, \bibinfo{author}{L.~Kador},
\newblock \bibinfo{journal}{Phys. Rev. Lett.} \bibinfo{volume}{62}
  (\bibinfo{year}{1989}) \bibinfo{pages}{2535--2538}.
\bibitem[{Zenhausern et~al.(1995)Zenhausern, Martin, and
  Wickramasinghe}]{zenhausern95}
\bibinfo{author}{F.~Zenhausern}, \bibinfo{author}{Y.~Martin},
  \bibinfo{author}{H.~K. Wickramasinghe},
\newblock \bibinfo{journal}{Science} \bibinfo{volume}{269}
  (\bibinfo{year}{1995}) \bibinfo{pages}{1083--1085}.
\bibitem[{Kador et~al.(1999)Kador, Latychevskaia, Renn, and Wild}]{kador99}
\bibinfo{author}{L.~Kador}, \bibinfo{author}{T.~Latychevskaia},
  \bibinfo{author}{A.~Renn}, \bibinfo{author}{U.~P. Wild},
\newblock \bibinfo{journal}{J. Chem. Phys.} \bibinfo{volume}{111}
  (\bibinfo{year}{1999}) \bibinfo{pages}{8755--8758}.
\bibitem[{Gerhardt et~al.(2007)Gerhardt, Wrigge, Bushev, Zumofen, Agio, Pfab,
  and Sandoghdar}]{gerhardt07a}
\bibinfo{author}{I.~Gerhardt}, \bibinfo{author}{G.~Wrigge},
  \bibinfo{author}{P.~Bushev}, \bibinfo{author}{G.~Zumofen},
  \bibinfo{author}{M.~Agio}, \bibinfo{author}{R.~Pfab},
  \bibinfo{author}{V.~Sandoghdar},
\newblock \bibinfo{journal}{Phys. Rev. Lett.} \bibinfo{volume}{98}
  (\bibinfo{year}{2007}) \bibinfo{pages}{033601}.
\bibitem[{Wrigge et~al.(2008)Wrigge, Gerhardt, Hwang, Zumofen, and
  Sandoghdar}]{wrigge08a}
\bibinfo{author}{G.~Wrigge}, \bibinfo{author}{I.~Gerhardt},
  \bibinfo{author}{J.~Hwang}, \bibinfo{author}{G.~Zumofen},
  \bibinfo{author}{V.~Sandoghdar},
\newblock \bibinfo{journal}{Nat. Phys.} \bibinfo{volume}{4}
  (\bibinfo{year}{2008}) \bibinfo{pages}{60--66}.
\bibitem[{Kukura et~al.(2010)Kukura, Celebrano, Renn, and
  Sandoghdar}]{kukura10}
\bibinfo{author}{P.~Kukura}, \bibinfo{author}{M.~Celebrano},
  \bibinfo{author}{A.~Renn}, \bibinfo{author}{V.~Sandoghdar},
\newblock \bibinfo{journal}{J. Phys. Chem. Lett.} \bibinfo{volume}{1}
  (\bibinfo{year}{2010}) \bibinfo{pages}{3323--3327}.
\bibitem[{Pototschnig et~al.(2011)Pototschnig, Chassagneux, Hwang, Zumofen,
  Renn, and Sandoghdar}]{pototschnig11}
\bibinfo{author}{M.~Pototschnig}, \bibinfo{author}{Y.~Chassagneux},
  \bibinfo{author}{J.~Hwang}, \bibinfo{author}{G.~Zumofen},
  \bibinfo{author}{A.~Renn}, \bibinfo{author}{V.~Sandoghdar},
\newblock \bibinfo{journal}{Phys. Rev. Lett.} \bibinfo{volume}{107}
  (\bibinfo{year}{2011}) \bibinfo{pages}{063001}.
\bibitem[{Celebrano et~al.(2011)Celebrano, Kukura, Renn, and
  Sandoghdar}]{celebrano11}
\bibinfo{author}{M.~Celebrano}, \bibinfo{author}{P.~Kukura},
  \bibinfo{author}{A.~Renn}, \bibinfo{author}{V.~Sandoghdar},
\newblock \bibinfo{journal}{Nat. Photon.} \bibinfo{volume}{5}
  (\bibinfo{year}{2011}) \bibinfo{pages}{95--98}.
\bibitem[{Zumofen et~al.(2008)Zumofen, Mojarad, Sandoghdar, and
  Agio}]{zumofen08}
\bibinfo{author}{G.~Zumofen}, \bibinfo{author}{N.~M. Mojarad},
  \bibinfo{author}{V.~Sandoghdar}, \bibinfo{author}{M.~Agio},
\newblock \bibinfo{journal}{Phys. Rev. Lett.} \bibinfo{volume}{101}
  (\bibinfo{year}{2008}) \bibinfo{pages}{180404}.
\bibitem[{Mojarad et~al.(2009)Mojarad, Zumofen, Sandoghdar, and
  Agio}]{mojarad09b}
\bibinfo{author}{N.~M. Mojarad}, \bibinfo{author}{G.~Zumofen},
  \bibinfo{author}{V.~Sandoghdar}, \bibinfo{author}{M.~Agio},
\newblock \bibinfo{journal}{J. Eur. Opt. Soc.: RP} \bibinfo{volume}{4}
  (\bibinfo{year}{2009}) \bibinfo{pages}{09014}.
\bibitem[{Greffet(2005)}]{greffet05}
\bibinfo{author}{J.-J. Greffet},
\newblock \bibinfo{journal}{Science} \bibinfo{volume}{308}
  (\bibinfo{year}{2005}) \bibinfo{pages}{1561--1563}.
\bibitem[{M\"uhlschlegel et~al.(2005)M\"uhlschlegel, Eisler, Martin, Hecht, and
  Pohl}]{muehlschlegel05}
\bibinfo{author}{P.~M\"uhlschlegel}, \bibinfo{author}{H.-J. Eisler},
  \bibinfo{author}{O.~J.~F. Martin}, \bibinfo{author}{B.~Hecht},
  \bibinfo{author}{D.~W. Pohl},
\newblock \bibinfo{journal}{Science} \bibinfo{volume}{308}
  (\bibinfo{year}{2005}) \bibinfo{pages}{1607--1609}.
\bibitem[{Novotny and van Hulst(2011)}]{novotny11}
\bibinfo{author}{L.~Novotny}, \bibinfo{author}{N.~van Hulst},
\newblock \bibinfo{journal}{Nat. Photon.} \bibinfo{volume}{5}
  (\bibinfo{year}{2011}) \bibinfo{pages}{83--90}.
\bibitem[{Anger et~al.(2006)Anger, Bharadwaj, and Novotny}]{anger06}
\bibinfo{author}{P.~Anger}, \bibinfo{author}{P.~Bharadwaj},
  \bibinfo{author}{L.~Novotny},
\newblock \bibinfo{journal}{Phys. Rev. Lett.} \bibinfo{volume}{96}
  (\bibinfo{year}{2006}) \bibinfo{pages}{113002}.
\bibitem[{K\"uhn et~al.(2006)K\"uhn, {H\aa kanson}, Rogobete, and
  Sandoghdar}]{kuehn06a}
\bibinfo{author}{S.~K\"uhn}, \bibinfo{author}{U.~{H\aa kanson}},
  \bibinfo{author}{L.~Rogobete}, \bibinfo{author}{V.~Sandoghdar},
\newblock \bibinfo{journal}{Phys. Rev. Lett.} \bibinfo{volume}{97}
  (\bibinfo{year}{2006}) \bibinfo{pages}{017402}.
\bibitem[{K\"uhn et~al.(2008)K\"uhn, Mori, Agio, and Sandoghdar}]{kuehn08}
\bibinfo{author}{S.~K\"uhn}, \bibinfo{author}{G.~Mori},
  \bibinfo{author}{M.~Agio}, \bibinfo{author}{V.~Sandoghdar},
\newblock \bibinfo{journal}{Mol. Phys.} \bibinfo{volume}{106}
  (\bibinfo{year}{2008}) \bibinfo{pages}{893--908}.
\bibitem[{Taminiau et~al.(2008)Taminiau, Stefani, Segerink, and van
  Hulst}]{taminiau08a}
\bibinfo{author}{T.~H. Taminiau}, \bibinfo{author}{F.~D. Stefani},
  \bibinfo{author}{F.~B. Segerink}, \bibinfo{author}{N.~F. van Hulst},
\newblock \bibinfo{journal}{Nat. Photon.} \bibinfo{volume}{2}
  (\bibinfo{year}{2008}) \bibinfo{pages}{234--237}.
\bibitem[{Curto et~al.(2010)Curto, Volpe, Taminiau, Kreuzer, Quidant, and van
  Hulst}]{curto10}
\bibinfo{author}{A.~G. Curto}, \bibinfo{author}{G.~Volpe},
  \bibinfo{author}{T.~H. Taminiau}, \bibinfo{author}{M.~P. Kreuzer},
  \bibinfo{author}{R.~Quidant}, \bibinfo{author}{N.~F. van Hulst},
\newblock \bibinfo{journal}{Science} \bibinfo{volume}{329}
  (\bibinfo{year}{2010}) \bibinfo{pages}{930--933}.
\bibitem[{Chen et~al.(2010)Chen, Sandoghdar, and Agio}]{chen10}
\bibinfo{author}{X.-W. Chen}, \bibinfo{author}{V.~Sandoghdar},
  \bibinfo{author}{M.~Agio},
\newblock \bibinfo{journal}{Opt. Express} \bibinfo{volume}{18}
  (\bibinfo{year}{2010}) \bibinfo{pages}{10878--10887}.
\bibitem[{Chang et~al.(2006)Chang, {S\o rensen}, Hemmer, and Lukin}]{chang06}
\bibinfo{author}{D.~E. Chang}, \bibinfo{author}{A.~S. {S\o rensen}},
  \bibinfo{author}{P.~R. Hemmer}, \bibinfo{author}{M.~D. Lukin},
\newblock \bibinfo{journal}{Phys. Rev. Lett.} \bibinfo{volume}{97}
  (\bibinfo{year}{2006}) \bibinfo{pages}{053002}.
\bibitem[{De~Angelis et~al.(2010)De~Angelis, Das, Candeloro, Patrini, Galli,
  Bek, Lazzarino, Maksymov, Liberale, Andreani, and Di~Fabrizio}]{deangelis10}
\bibinfo{author}{F.~De~Angelis}, \bibinfo{author}{G.~Das},
  \bibinfo{author}{P.~Candeloro}, \bibinfo{author}{M.~Patrini},
  \bibinfo{author}{M.~Galli}, \bibinfo{author}{A.~Bek},
  \bibinfo{author}{M.~Lazzarino}, \bibinfo{author}{I.~Maksymov},
  \bibinfo{author}{C.~Liberale}, \bibinfo{author}{L.~C. Andreani},
  \bibinfo{author}{E.~Di~Fabrizio},
\newblock \bibinfo{journal}{Nat. Nano.} \bibinfo{volume}{5}
  (\bibinfo{year}{2010}) \bibinfo{pages}{67--72}.
\bibitem[{Takahara et~al.(1997)Takahara, Yamagishi, Taki, Morimoto, and
  Kobayashi}]{takahara97}
\bibinfo{author}{J.~Takahara}, \bibinfo{author}{S.~Yamagishi},
  \bibinfo{author}{H.~Taki}, \bibinfo{author}{A.~Morimoto},
  \bibinfo{author}{T.~Kobayashi},
\newblock \bibinfo{journal}{Opt. Lett.} \bibinfo{volume}{22}
  (\bibinfo{year}{1997}) \bibinfo{pages}{475--477}.
\bibitem[{Babadjanyan et~al.(2000)Babadjanyan, Margaryan, and
  Nerkararyan}]{babadjanyan00}
\bibinfo{author}{A.~J. Babadjanyan}, \bibinfo{author}{N.~L. Margaryan},
  \bibinfo{author}{K.~V. Nerkararyan},
\newblock \bibinfo{journal}{J. Appl. Phys.} \bibinfo{volume}{87}
  (\bibinfo{year}{2000}) \bibinfo{pages}{3785--3788}.
\bibitem[{Stockman(2004)}]{stockman04}
\bibinfo{author}{M.~I. Stockman},
\newblock \bibinfo{journal}{Phys. Rev. Lett.} \bibinfo{volume}{93}
  (\bibinfo{year}{2004}) \bibinfo{pages}{137404}.
\bibitem[{Gramotnev and Bozhevolnyi(2010)}]{gramotnev10}
\bibinfo{author}{D.~K. Gramotnev}, \bibinfo{author}{S.~I. Bozhevolnyi},
\newblock \bibinfo{journal}{Nat. Photon.} \bibinfo{volume}{4}
  (\bibinfo{year}{2010}) \bibinfo{pages}{83--91}.
\bibitem[{Bohren and Huffman(1983)}]{bohren83b}
\bibinfo{author}{C.~F. Bohren}, \bibinfo{author}{D.~R. Huffman},
  \bibinfo{title}{Absorption and Scattering of Light by Small Particles},
  \bibinfo{publisher}{John Wiley \& Sons}, \bibinfo{address}{New York},
  \bibinfo{year}{1983}.
\bibitem[{Lock et~al.(1995)Lock, Hodges, and Gouesbet}]{lock95b}
\bibinfo{author}{J.~A. Lock}, \bibinfo{author}{J.~T. Hodges},
  \bibinfo{author}{G.~Gouesbet},
\newblock \bibinfo{journal}{J. Opt. Soc. Am. A} \bibinfo{volume}{12}
  (\bibinfo{year}{1995}) \bibinfo{pages}{2708--2715}.
\bibitem[{Jackson(1999)}]{jackson99}
\bibinfo{author}{J.~D. Jackson}, \bibinfo{title}{Classical Electrodynamics},
  \bibinfo{publisher}{John Wiley \& Sons}, \bibinfo{address}{New York},
  \bibinfo{edition}{third} edition, \bibinfo{year}{1999}.
\bibitem[{Allen and Eberly(1975)}]{allen75}
\bibinfo{author}{L.~Allen}, \bibinfo{author}{J.~Eberly},
  \bibinfo{title}{Optical resonance and two-level atoms},
  \bibinfo{publisher}{Dover}, \bibinfo{address}{New York},
  \bibinfo{year}{1975}.
\bibitem[{Bassett(1986)}]{bassett86}
\bibinfo{author}{I.~M. Bassett},
\newblock \bibinfo{journal}{Optica Acta} \bibinfo{volume}{33}
  (\bibinfo{year}{1986}) \bibinfo{pages}{279--286}.
\bibitem[{Zumofen et~al.(2009)Zumofen, Mojarad, and Agio}]{zumofen09}
\bibinfo{author}{G.~Zumofen}, \bibinfo{author}{N.~M. Mojarad},
  \bibinfo{author}{M.~Agio},
\newblock \bibinfo{journal}{N. Cimento C} \bibinfo{volume}{31}
  (\bibinfo{year}{2009}) \bibinfo{pages}{475--485}.
\bibitem[{Aljunid et~al.(2009)Aljunid, Tey, Chng, Liew, Maslennikov, Scarani,
  and Kurtsiefer}]{aljunid09}
\bibinfo{author}{S.~A. Aljunid}, \bibinfo{author}{M.~K. Tey},
  \bibinfo{author}{B.~Chng}, \bibinfo{author}{T.~Liew},
  \bibinfo{author}{G.~Maslennikov}, \bibinfo{author}{V.~Scarani},
  \bibinfo{author}{C.~Kurtsiefer},
\newblock \bibinfo{journal}{Phys. Rev. Lett.} \bibinfo{volume}{103}
  (\bibinfo{year}{2009}) \bibinfo{pages}{153601}.
\bibitem[{Agio et~al.(2009)Agio, Zumofen, Mojarad, and Sandoghdar}]{agio09}
\bibinfo{author}{M.~Agio}, \bibinfo{author}{G.~Zumofen}, \bibinfo{author}{N.~M.
  Mojarad}, \bibinfo{author}{V.~Sandoghdar},
\newblock in: \bibinfo{editor}{S.~Kawata}, \bibinfo{editor}{V.~M. Shalaev},
  \bibinfo{editor}{D.~P. Tsai} (Eds.), \bibinfo{booktitle}{Plasmonics:
  Nanoimaging, Nanofabrication, and their Applications V}, volume
  \bibinfo{volume}{7395}, \bibinfo{publisher}{Proc. SPIE},
  \bibinfo{year}{2009}, p. \bibinfo{pages}{739512}.
\bibitem[{Gordon(2009)}]{gordon09}
\bibinfo{author}{R.~Gordon},
\newblock \bibinfo{journal}{Opt. Express} \bibinfo{volume}{17}
  (\bibinfo{year}{2009}) \bibinfo{pages}{18621--18629}.
\bibitem[{Vogel and Gramotnev(2007)}]{vogel07}
\bibinfo{author}{M.~W. Vogel}, \bibinfo{author}{D.~K. Gramotnev},
\newblock \bibinfo{journal}{Phys. Lett. A} \bibinfo{volume}{363}
  (\bibinfo{year}{2007}) \bibinfo{pages}{507--511}.
\bibitem[{Gramotnev et~al.(2008)Gramotnev, Vogel, and Stockman}]{gramotnev08}
\bibinfo{author}{D.~K. Gramotnev}, \bibinfo{author}{M.~W. Vogel},
  \bibinfo{author}{M.~I. Stockman},
\newblock \bibinfo{journal}{J. Appl. Phys.} \bibinfo{volume}{104}
  (\bibinfo{year}{2008}) \bibinfo{pages}{034311}.
\bibitem[{Rogobete et~al.(2007)Rogobete, Kaminski, Agio, and
  Sandoghdar}]{rogobete07a}
\bibinfo{author}{L.~Rogobete}, \bibinfo{author}{F.~Kaminski},
  \bibinfo{author}{M.~Agio}, \bibinfo{author}{V.~Sandoghdar},
\newblock \bibinfo{journal}{Opt. Lett.} \bibinfo{volume}{32}
  (\bibinfo{year}{2007}) \bibinfo{pages}{1623--1625}.
\bibitem[{Taflove and Hagness(2005)}]{taflove05}
\bibinfo{author}{A.~Taflove}, \bibinfo{author}{S.~C. Hagness},
  \bibinfo{title}{Computational Electrodynamics: The Finite-Difference
  Time-Domain Method}, \bibinfo{publisher}{Artech House},
  \bibinfo{address}{Norwood, MA}, \bibinfo{edition}{third} edition,
  \bibinfo{year}{2005}.
\bibitem[{Wokaun et~al.(1982)Wokaun, Gordon, and Liao}]{wokaun82}
\bibinfo{author}{A.~Wokaun}, \bibinfo{author}{J.~P. Gordon},
  \bibinfo{author}{P.~F. Liao},
\newblock \bibinfo{journal}{Phys. Rev. Lett.} \bibinfo{volume}{48}
  (\bibinfo{year}{1982}) \bibinfo{pages}{957--960}.
\bibitem[{Ashcroft and Mermin(1976)}]{ashcroft76}
\bibinfo{author}{N.~W. Ashcroft}, \bibinfo{author}{N.~D. Mermin},
  \bibinfo{title}{Solid State Physics}, \bibinfo{publisher}{Saunders College
  Publishing}, \bibinfo{address}{Fort Worth}, \bibinfo{year}{1976}.
\bibitem[{Ambrose et~al.(1994)Ambrose, Goodwin, Keller, and Martin}]{ambrose94}
\bibinfo{author}{W.~P. Ambrose}, \bibinfo{author}{P.~M. Goodwin},
  \bibinfo{author}{R.~A. Keller}, \bibinfo{author}{J.~C. Martin},
\newblock \bibinfo{journal}{Science} \bibinfo{volume}{265}
  (\bibinfo{year}{1994}) \bibinfo{pages}{364--367}.
\bibitem[{Bian et~al.(1995)Bian, Dunn, Xie, and Leung}]{bian95}
\bibinfo{author}{R.~X. Bian}, \bibinfo{author}{R.~C. Dunn},
  \bibinfo{author}{X.~S. Xie}, \bibinfo{author}{P.~T. Leung},
\newblock \bibinfo{journal}{Phys. Rev. Lett.} \bibinfo{volume}{75}
  (\bibinfo{year}{1995}) \bibinfo{pages}{4772--4775}.
\bibitem[{Ropers et~al.(2007)Ropers, Neacsu, Elsaesser, Albrecht, Raschke, and
  Lienau}]{ropers07}
\bibinfo{author}{C.~Ropers}, \bibinfo{author}{C.~C. Neacsu},
  \bibinfo{author}{T.~Elsaesser}, \bibinfo{author}{M.~Albrecht},
  \bibinfo{author}{M.~B. Raschke}, \bibinfo{author}{C.~Lienau},
\newblock \bibinfo{journal}{Nano Lett.} \bibinfo{volume}{7}
  (\bibinfo{year}{2007}) \bibinfo{pages}{2784--2788}.
\bibitem[{Neacsu et~al.(2010)Neacsu, Berweger, Olmon, Saraf, Ropers, and
  Raschke}]{neacsu10}
\bibinfo{author}{C.~Neacsu}, \bibinfo{author}{S.~Berweger},
  \bibinfo{author}{R.~Olmon}, \bibinfo{author}{L.~Saraf},
  \bibinfo{author}{C.~Ropers}, \bibinfo{author}{M.~Raschke},
\newblock \bibinfo{journal}{Nano Lett.} \bibinfo{volume}{10}
  (\bibinfo{year}{2010}) \bibinfo{pages}{592--596}.
\bibitem[{Sadiq et~al.(2011)Sadiq, Shirdel, Lee, Selishcheva, Park, and
  Lienau}]{sadiq11}
\bibinfo{author}{D.~Sadiq}, \bibinfo{author}{J.~Shirdel},
  \bibinfo{author}{J.~Lee}, \bibinfo{author}{E.~Selishcheva},
  \bibinfo{author}{N.~Park}, \bibinfo{author}{C.~Lienau},
\newblock \bibinfo{journal}{Nano Lett.} \bibinfo{volume}{11}
  (\bibinfo{year}{2011}) \bibinfo{pages}{1609--1613}.
\bibitem[{Ridolfo et~al.(2010)Ridolfo, Di~Stefano, Fina, Saija, and
  Savasta}]{ridolfo10}
\bibinfo{author}{A.~Ridolfo}, \bibinfo{author}{O.~Di~Stefano},
  \bibinfo{author}{N.~Fina}, \bibinfo{author}{R.~Saija},
  \bibinfo{author}{S.~Savasta},
\newblock \bibinfo{journal}{Phys. Rev. Lett.} \bibinfo{volume}{105}
  (\bibinfo{year}{2010}) \bibinfo{pages}{263601}.
\bibitem[{Alaverdyan et~al.(2011)Alaverdyan, Vamivakas, Barnes, Lebouteiller,
  Hare, and Atat\"{u}re}]{alaverdyan11}
\bibinfo{author}{Y.~Alaverdyan}, \bibinfo{author}{N.~Vamivakas},
  \bibinfo{author}{J.~Barnes}, \bibinfo{author}{C.~Lebouteiller},
  \bibinfo{author}{J.~Hare}, \bibinfo{author}{M.~Atat\"{u}re},
\newblock \bibinfo{journal}{Opt. Express} \bibinfo{volume}{19}
  (\bibinfo{year}{2011}) \bibinfo{pages}{18175--18181}.
\bibitem[{S\'anchez et~al.(1999)S\'anchez, Novotny, and Xie}]{sanchez99}
\bibinfo{author}{E.~J. S\'anchez}, \bibinfo{author}{L.~Novotny},
  \bibinfo{author}{X.~S. Xie},
\newblock \bibinfo{journal}{Phys. Rev. Lett.} \bibinfo{volume}{82}
  (\bibinfo{year}{1999}) \bibinfo{pages}{4014--4017}.
\bibitem[{Guenther et~al.(2002)Guenther, Lienau, Elsaesser, Glanemann, Axt,
  Kuhn, Eshlaghi, and Wieck}]{guenther02}
\bibinfo{author}{T.~Guenther}, \bibinfo{author}{C.~Lienau},
  \bibinfo{author}{T.~Elsaesser}, \bibinfo{author}{M.~Glanemann},
  \bibinfo{author}{V.~M. Axt}, \bibinfo{author}{T.~Kuhn},
  \bibinfo{author}{S.~Eshlaghi}, \bibinfo{author}{A.~D. Wieck},
\newblock \bibinfo{journal}{Phys. Rev. Lett.} \bibinfo{volume}{89}
  (\bibinfo{year}{2002}) \bibinfo{pages}{057401}.
\bibitem[{Ichimura et~al.(2004)Ichimura, Hayazawa, Hashimoto, Inouye, and
  Kawata}]{ichimura04}
\bibinfo{author}{T.~Ichimura}, \bibinfo{author}{N.~Hayazawa},
  \bibinfo{author}{M.~Hashimoto}, \bibinfo{author}{Y.~Inouye},
  \bibinfo{author}{S.~Kawata},
\newblock \bibinfo{journal}{Phys. Rev. Lett.} \bibinfo{volume}{92}
  (\bibinfo{year}{2004}) \bibinfo{pages}{220801}.
\bibitem[{Celebrano et~al.(2009)Celebrano, Biagioni, Zavelani-Rossi, Polli,
  Labardi, Allegrini, Finazzi, Du\`{o}, and Cerullo}]{celebrano09}
\bibinfo{author}{M.~Celebrano}, \bibinfo{author}{P.~Biagioni},
  \bibinfo{author}{M.~Zavelani-Rossi}, \bibinfo{author}{D.~Polli},
  \bibinfo{author}{M.~Labardi}, \bibinfo{author}{M.~Allegrini},
  \bibinfo{author}{M.~Finazzi}, \bibinfo{author}{L.~Du\`{o}},
  \bibinfo{author}{G.~Cerullo},
\newblock \bibinfo{journal}{Rev. Sci. Instr.} \bibinfo{volume}{80}
  (\bibinfo{year}{2009}) \bibinfo{pages}{033704}.
\bibitem[{Berweger et~al.(2011)Berweger, Atkin, Xu, Olmon, and
  Raschke}]{berweger11}
\bibinfo{author}{S.~Berweger}, \bibinfo{author}{J.~M. Atkin},
  \bibinfo{author}{X.~G. Xu}, \bibinfo{author}{R.~L. Olmon},
  \bibinfo{author}{M.~B. Raschke},
\newblock \bibinfo{journal}{Nano Lett.} \bibinfo{volume}{11}
  (\bibinfo{year}{2011}) \bibinfo{pages}{4309--4313}.
\bibitem[{Brinks et~al.(2010)Brinks, Stefani, Kulzer, Hildner, Taminiau,
  Avlasevich, Mullen, and van Hulst}]{brinks10}
\bibinfo{author}{D.~Brinks}, \bibinfo{author}{F.~D. Stefani},
  \bibinfo{author}{F.~Kulzer}, \bibinfo{author}{R.~Hildner},
  \bibinfo{author}{T.~H. Taminiau}, \bibinfo{author}{Y.~Avlasevich},
  \bibinfo{author}{K.~Mullen}, \bibinfo{author}{N.~F. van Hulst},
\newblock \bibinfo{journal}{Nature} \bibinfo{volume}{465}
  (\bibinfo{year}{2010}) \bibinfo{pages}{905--908}.
\bibitem[{Mukamel(2000)}]{mukamel00}
\bibinfo{author}{S.~Mukamel},
\newblock \bibinfo{journal}{Annu. Rev. Phys. Chem.} \bibinfo{volume}{51}
  (\bibinfo{year}{2000}) \bibinfo{pages}{691--729}.
\bibitem[{Aeschlimann et~al.(2011)Aeschlimann, Brixner, Fischer, Kramer,
  Melchior, Pfeiffer, Schneider, Str√ºber, Tuchscherer, and
  Voronine}]{aeschlimann11}
\bibinfo{author}{M.~Aeschlimann}, \bibinfo{author}{T.~Brixner},
  \bibinfo{author}{A.~Fischer}, \bibinfo{author}{C.~Kramer},
  \bibinfo{author}{P.~Melchior}, \bibinfo{author}{W.~Pfeiffer},
  \bibinfo{author}{C.~Schneider}, \bibinfo{author}{C.~Str√ºber},
  \bibinfo{author}{P.~Tuchscherer}, \bibinfo{author}{D.~V. Voronine},
\newblock \bibinfo{journal}{Science}  (\bibinfo{year}{2011}).
\bibitem[{Hillenbrand et~al.(2002)Hillenbrand, Taubner, and
  Keilmann}]{hillenbrand02}
\bibinfo{author}{R.~Hillenbrand}, \bibinfo{author}{T.~Taubner},
  \bibinfo{author}{F.~Keilmann},
\newblock \bibinfo{journal}{Nature} \bibinfo{volume}{418}
  (\bibinfo{year}{2002}) \bibinfo{pages}{159--162}.
\bibitem[{Maier et~al.(2006)Maier, Andrews, Mart\'in-Moreno, and
  Garc\'ia-Vidal}]{maier06a}
\bibinfo{author}{S.~A. Maier}, \bibinfo{author}{S.~R. Andrews},
  \bibinfo{author}{L.~Mart\'in-Moreno}, \bibinfo{author}{F.~J. Garc\'ia-Vidal},
\newblock \bibinfo{journal}{Phys. Rev. Lett.} \bibinfo{volume}{97}
  (\bibinfo{year}{2006}) \bibinfo{pages}{176805}.
\bibitem[{Goubau(1950)}]{goubau50}
\bibinfo{author}{G.~Goubau},
\newblock \bibinfo{journal}{J. Appl. Phys.} \bibinfo{volume}{21}
  (\bibinfo{year}{1950}) \bibinfo{pages}{1119--1128}.

\end{thebibliography}








\end{document}